# An Optimal Load Balanced Resource Allocation Scheme for Heterogeneous Wireless Networks based on Big Data Technology

Abbas Mirzaei
Department of Information, Communications & Security Technology, Malek Ashtar University of Technology, Tehran, Iran

Morteza Barari
Department of Information, Communications & Security Technology, Malek Ashtar University of Technology, Tehran, Iran

Houman Zarrabi
Integrated Network Management Group
Iran Télécommunication Research Center (ITRC),
Tehran, Iran

*Abstract*—An important issue in heterogeneous wireless networks is how to optimally utilize various radio resources. While many methods have been proposed for managing radio resources in each network, these methods are not suitable for heterogeneous wireless networks. In this study, a new management method is proposed which provides acceptable service quality and roaming rate, reduces the cost of the service, and utilizes big data technology for its operation. In our proposed scheme, by considering various parameters such as the type of the service, the information related to the location of the user, the movement direction of the user, the cost of the service, and a number of other statistical measures, the most suitable technology for radio access will be selected. It is expected that besides improving the decision making accuracy in selecting the radio access technology and the balanced distribution of network resources, the proposed method provides lower roaming and lower probability of stopping roaming requests entering the network. By considering the various service classes and various quality of service requirements regarding delay, vibration and so on, this can be useful in optimal implementation of heterogeneous wireless networks.

I. INTRODUCTION

Now-a-days, various technologies with various goals are proposed for wireless access. For instance, local wireless networks using the IEEE 802.11 standard, wireless urban networks using IEEE 802.16 standard, and cellular mobile networks (4G); considering that these networks vary based on bandwidth, the type of access to media, and the security they provide for the users [1], [2]. However, heterogeneous combinatory networks provide the possibility of optimizing access everywhere and under various conditions by maximizing the utilization of the facilities of these technologies. Previously, decision making for connecting to the network was up to the network and roaming only occurred when the signal was weak. However, in heterogeneous wireless networks, when the possibility of accessing the internet for the users is possible through more than one network, the users select the best type of access for their connections in real time.

In other words, before the introduction of next generation heterogeneous wireless networks, selecting the radio station was done by the network and the only criterion was the strength of the received signal [3]. However, the fact that the user can select not only from stations with the same technology, but also from heterogeneous technologies with various features is a new possibility which creates new challenges [7], [8]. While usually the high bandwidth of local wireless networks such as 802.11 is very important for data transmission, due to the small coverage area, these networks are only available in hotspots such as hotels and libraries, forcing the moving user to return to the mobile network. While WiMAX networks provide acceptable bandwidth and quality of service, the significant distance between the stations and the different frequency and technology used for transmitting information increase the power consumption of the user utilizing this technology and if the battery of the user's device is very low, they can connect to the mobile network without losing the connection in order to consume less power [11]-[15]. Moreover, using ad hoc networks and the Bluetooth technology, they can connect to a modem and keep the connection with a very low power consumption. In heterogeneous access network, all the advantages of each one of these networks is effectively available in a combinatory network and the main goal of heterogeneous wireless networks is the realization of the concept of 'Always Best Connected'. In other words, users select the best option among the provided service based on their conditions [9], [10]; an option which is proportionate to their preferences and requirements so that they can utilize the best facilities. Accordingly, the user will pay the lowest price for the best possible quality of service. There are various decision making parameters for variable conditions of each running application and the conditions of the networks and the device of the user. If the system is to consider physical parameters, such as quality of signal and the bit error, as well as the requirements of the user in selecting networks with lower cost and lower congestion, the decision making problem will be a multi-dimensional problem with conflicting objectives. Furthermore, selecting the best network must be based on policy, priority, and the sensitivity of these criteria as well as the effects it has on more general goals such as the load distribution. Another issue in heterogeneous wireless networks is the nature of distributed access. In heterogeneous networks where even selecting the various access technologies of the network is up to the users, we have to meet the general goals of the network such as balanced load or optimized service, meaning that we are faced with a distributed optimization problem. The balanced allocation of load among the networks and the stations will not only prevent congestion but also lead to balanced utilization of network resources. The users themselves are also willing to select the best network regarding load and congestion so that they can receive the best quality of service. However, the user will not consider only one criterion; they care about other local criteria such as the degree of power consumption, the quality of service, the movement of the user, and so on. Therefore, the outcome of users' decisions and strategies and their effects of the general goals of the network operators, such as balanced load, must be considered and modelled. Overall, the decisions the user makes for receiving the best quality with the lowest cost lead to the formation of general network goals. The general network goals such as balanced load, increasing the revenue of operators, and increasing the overall productivity of the network must be considered along with the decision goals of the user. Hence, modeling and the distributed optimization analysis mentioned above are very important in next generation wireless access networks [5], [6].

In this study, we will discuss the problem of load allocation in next generation mobile networks based on data analysis methods such as big data method and roaming among mobile networks and WLAN. The rest of this paper is organized as follows. In Section 2, we review the research literature described in this paper. In Section 3, the importance of big data is stated for the proposed model. In Section 4, the proposed model is described in details. The evaluation results have been discussed in Section 5. Finally in Section 6, the conclusion and future work are expressed.

## II. RELATED WORK

The main problem in the study is allocating load to femtocell users and roaming from the mobile network to WLAN. Hence, at first the user's device connects to the mobile network and if the mobile network is not able to provide the selected service to the user's device for any reason whatsoever, the process of roaming to WLAN initiates [12]. On the other hand, when for any reason such as the occurrence of an accident or event, the number of subscribers in that location is too large, using big data, it is intelligent enough to manage roaming from one network to another considering the network resources and the event that happened in that location. In this study, the roaming from mobile to WLAN occurs if necessary. In other words, if it can lead to improved quality, it will be considered necessary. Therefore, ideally this operation must occur when we are sure of improved quality. The important issue in heterogeneous wireless networks is how to utilize various radio resources. While there are many methods for managing radio resources in each network separately, these methods are not suitable for heterogeneous wireless networks [17], [18]. Hence, a new concept has emerged as the continuous management of radio resources [4]. The continuous management of radio resources refers to a set of function for effective and coordinated utilization of radio resources in heterogeneous networks. Methods for selecting RAT, which includes algorithms for selecting Rat at the beginning of a session as well as selecting RAT when shifting running communication from one RAT to another Rat (vertical roaming), are the main element of managing radio resources in heterogeneous wireless networks. Having reviewed the studies related to managing resources in heterogeneous wireless networks, the following methods can be extracted: first, methods for designing the optimal target function, and second, the theoretical methods which pursue goals other than the simple optimization of a function. The fuzzy method and the game theory are instances of the second set of methods which analyze problems such as competition, collaboration and optimization for a number of selfish decision makers and extract an optimal model for the entire system. Various studies utilize fuzzy logic for decision makings related to inter-network roaming and resource allocation. The advantage of these methods is that they carry out the decision making without requiring detailed and explicit data. Decision making for selecting the optimal network is a multi-dimensional decision making problem and it is a type of approximate inference. Table 1 illustrates an example of fuzzy decision rules.

TABLE I. FUZZIFIED PARAMETERS

| |
|---|
| C={Economic, Normal, Expensive} |
| B={Poor, Med, Good} |
| RSS={Low, Normal, High} |
| U={Insensitive, Ordinary, High QoS} |
| D={Low, Med, High} |

The game theory models the interaction among the players for reaching common resources. Hence, it is suitable for modeling the management of radio resources. This study classifies the cells into concentric rings with the radius of where . The distance between each two rings is called a zone. Then, for each status of the system, which is denoted using the number of users in each zone (a non-collaborating game is designed. The profit function is based on the throughput in each zone which depends on the total number of mobile users in it and the number of those who use WiMAX or 3G networks. In a non-collaborating game, the optimal solution is the Nash equilibrium where the strategy of the players dictates that no player receives more profit from changing its strategy unilaterally (while the strategies of others remain constant). Therefore, the selected strategy of each player is the optimal solution for the strategies of other players [20]-[22].

The study models the issue of selecting a network for users facing competitive operators using non-collaborating games among the users. This game belongs to a class of games called 'swarm game'. The equilibrium points in this game are calculated and their quality will be evaluated based on the nominal throughput of the networks and their interference level. In all the above studies, the game is carried out based on geographical division of the location into zones or areas (each one based on a model) and then modeling the congestion level in that zone or area, which is a function of the number of users in each zone. The Nash equilibrium point in game models provides the best possible way of distributing the users among these geographical zones. The issue is that such a model is suitable for pre-design stages; for instance, when selecting the locations for radio stations and they do not discuss the willingness to improve equilibrium points for preventing congestion and besides modeling bandwidth, other important requirements of the users such as power consumption exerted by each network on the users or the difference between the nominal bandwidth of each network, which is very important in decision making, are not considered. In other words, the contribution of users in creating congestion or their need for bandwidth is considered equal, which is not true for modern data networks.

### III. IMPORTANCE OF BIG DATA IN THE PROPOSED MODEL

In the selected architecture of the proposed scheme, the information related to the users and the conditions of the network reach the network's core layer through uplink and through a network defined process, the decision making based on data analysis in this level of network will be carried out. Naturally, a part of the network's core is responsible for data analysis and decision making, known as the main switching unit. Based on the above figure, this switching unit in third generation networks is the SGSN and in next generation networks such as 4G and NGMN, it will be SGW and MME switching units. Therefore, in the first stage, a capacity for storing the user data with a large volume in the network core must be considered; in the next generation mobile networks which work based on IP, these data will never reach the switching unit. However, when we are willing to use data analysis techniques for network defined decision making, we will naturally need the uplink information. In case the current registers available in the network core are not able to store this volume of data, we have to consider a separate register in the network which provides the required capacity and is also able to quickly connect to the switching center. In this case, we have to add a section to the network core where data analysis based on big data technology is carried out. In Fig. 1, we will discuss the proposed idea in more details.

In this paper, by using the methods of huge data analysis, our purpose will be obtaining to the self-optimization limitation in multi carrier networks. After surveying the presented models, it will be tried to present solutions for some present challenges in the mobile networks field of new generation. We will pay attention the suggested model. In this model, we have tried to cover the present defects in past designs. General framework of this design has been formed from different parts that have linked like block diagram to each other. These parts can be interpreted in relation with a self-organized system with the technology of huge data. But, in continuance, we will pay attention for complete surveying of suggested idea in the framework of this diagram block.

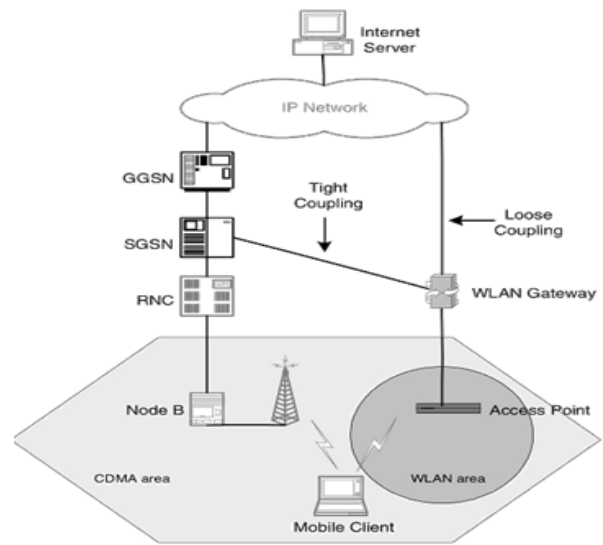

Fig. 1. The Importance of Big Data in the Proposed Model.

It should be noticed that because our work field is mobile networks of new generation, so, it is natural that also our way of receiving information will be the way of Up Link of network from linked things to Femtocells and also different users till the core network [3]. All of received information will be used for analyzing of determination of rate accuracy of dedicated sources till we have suitable control on internal and external interference rate. Next, we will survey physical model of Femtocell layer. In this study, we have tried to present an approximate model of optimization relation that has been used in the layer structure of multi carrier network. The next step in self-optimization networks, by certain approaches that there is measuring management and is able for learning, will be introduced based on past and present observations. This is the method of self-organized networks that are not limited by pre-introduced algorithm and in this pattern, one part of the network are able to arrange any sudden conditions. In this design, we have observed a capability that has been introduced according with this new approach till the nodes of a network also can have the ability of teaching. With performing the algorithms of machine learning, they will be able for exchanging information and teaching each other that will accelerate the process of self-teaching and will cause to faster convergence. Movement towards the changes of fifth generation not only there is in accessible radio network but also there is the section of core of network that its new approach makes possible the designing needed network for presenting services accordance with the users and increasing sets. The trend is like this that separating hardware and software and the movement of functions is one second. In order to extending the presented model to real conditions, it is necessary to observe the network obligations in different conditions and different periods. As we observed in surveying past works, the affairs that have been posed in the field of optimizing mobile networks and presenting a self-optimization model, it performs the intelligence of network only through analyzing KPI telecommunication efficiency parameters, and they neglect effective non-telecommunication parameters in distributing network traffic and transferring network bar, and based on, the destructive outcomes that there are in mobile networks due to environmental and social factors, will not be considered in this kind of designs. In order to implement the plan to self-optimization based on data analysis, we divide the self-parameters into two general following categories. Our purpose is reaching to the self-optimization pattern that in this design one part of the network, based on supervision on the parameters of these two groups, can predict the conditions of is future and to estimate the rate of its needed sources according network bar in every part of the network. The pattern that we will present from self-optimization will contain two parts of three parts in performance of self-organized systems.

IV. PROPOSED MODEL

*A. User Mobility Model*

In this study, a movement model based on probabilities according to [15] is considered for the user. Accordingly, the probability of the user exiting the hotspot in their session is denoted by and the probability that they enter a hotspot is denoted by Moreover, the probability that the user is a vehicular user (with a lot of movement) is shown as and the probability of the user being a non-vehicular user is shown as.

Considering low values for identifies a hotspot scenario for a company where users will stay in the hotspot area (their own company) for a very long time. In contrast, high values for indicate a public area covered by WLAN (e.g. airports) where users are moving in and out all the time.

Under equilibrium conditions, the average number of users exiting the hotspot and the average number of users entering this area will be equal [16]. Therefore, we will have:

The values of and play a significant role in analyzing the VHO process in HWN environments. For vehicular users we will have a very high velocity, so we assume that the probability of them changing their location from outside a hotspot to inside a hotspot is. Consequently, we will have:

Using conditional probability and (5), the probability that the user is non-vehicular and the probability of them changing their location will be calculated as follows:

The main characteristic of this method is considering various parameters such as the type of user services, the type of user's movement, the current location and the future location of the user in the decision making process. Using this information, JRRM can make optimal decisions in order to prevent repetitive and unnecessary handovers, reducing the cost of services for the operators, which will be very important in implementing heterogeneous wireless networks. Another characteristic of this method is ensuring the quality of services for the users and maintaining the stability of user services.

*B. RAT Selection Algorithm based on Location and User Mobility*

In this JRRM algorithm, the necessary decisions for selecting an appropriate RAT are made according to the information it has gathered. RAT selection algorithm consists of three stages, as it is shown is the Table 2. JRRM entity receives the request for a new call.

JRRM requests for the location-related information and user mobility from the units of location register and location predictor of the user. According to the gathered information by the JRRM unit and the rules that come in the following, the most appropriate RAT for the session will be selected.

*1)* If the user is outside the hot-spot area, resource management will be done in LTE by the available radio resource management unit.

*2)* Otherwise, for an automotive user that is in the hot-spot area, LTE will be selected for different resource management so that the repetitive VHO which may happen due to the frequent automotive user mobility is prevented from.

*3)* The RAT service type will be selected for a non-automotive user that is in the hot-spot area.

If the condition (service type=non-immediate service) is on, IEEE 802.11 will be selected for the user due to the high-bandwidth and low-service cost. Otherwise, if the condition (service type=immediate service) is on, an appropriate RAT will be selected according to the relevant information in regards to user location prediction.

As in case it is predicted that the user has left the hot-spot area, IEEE 802.11 will be selected for him/her (so that service resistibility is guaranteed) otherwise LTE will be an appropriate option for the user so as to prevent from an unnecessary VHO.

TABLE II. PROPERTIES OF THE VHO ALGORITHM

|  | **Big data analysis for location prediction unit** | **Big data analysis for location registration unit** |  |
|---|---|---|---|
| User terminal Type of service | JPRM Type of user service Type of user movement The current and future locations of the user |  |  |
|  | User location |  | Outside hotspot |
|  | Inside hotspot |  |  |
|  | Type of the user through big data analysis |  | Vehicular user |
|  | Non-vehicular user |  |  |
| Non-real-time | Type of user service |  |  |
|  | Real-time |  |  |
|  | Future location of the user through big data analysis |  |  |
| 802.11 RRM | Staying in hotspot | Leaving the hotspot | LTE |

Continuing the explanation of this flowchart, in case the WLAN network finally is considered as the target network, the capacity of this network must be considered. In case the capacity of this network isn't providing our needs for the entrance of new users and using the WLAN network resources, we should inevitably select the cellular network as our target network.

In this JRRM algorithm, the necessary decisions for selecting an appropriate RAT are made according to the information it has gathered. RAT selection algorithm consists of three stages. JRRM entity receives the request for a new call.

JRRM requests for the location-related information and user mobility from the units of location register and location predictor of the user. According to the gathered information by the JRRM unit and the rules that come in the following, the most appropriate RAT for the session will be selected.

- If the user is outside the hot-spot area, resource management will be done in LTE by the available radio resource management unit.

- Otherwise, for an automotive user that is in the hot-spot area, LTE will be selected for different resource management so that the repetitive VHO which may happen due to the frequent automotive user mobility is prevented from.

- The RAT service type will be selected for a non-automotive user that is in the hot-spot area.

If the condition (service type=non-immediate service) is on, IEEE 802.11 will be selected for the user due to the high-bandwidth and low-service cost. Otherwise, if the condition (service type=immediate service) is on, an appropriate RAT will be selected according to the relevant information in regards to user location prediction.

As in case it is predicted that the user has left the hot-spot area, IEEE 802.11 will be selected for him/her (so that service resistibility is guaranteed) otherwise LTE will be an appropriate option for the user so as to prevent from an unnecessary VHO.

Continuing the explanation of this flowchart, in case the WLAN network finally is considered as the target network, the capacity of this network must be considered. In case the capacity of this network isn't providing our needs for the entrance of new users and using the WLAN network resources, we should inevitably select the cellular network as our target network [19].

*C. Multi-Dimensional Decision-Making*

The proposed pattern that was presented in this paper is based on a set of multi-dimensional decision-making procedures. In this pattern, one of the packets of the network is used based on three basic parameters of the service type provided, the expense of service provision as well as user mobility type and of course in the assumed decision-makings, the number of vertical inter-systemic handovers are supposed to decline.

Each one of the aforementioned indices can to different efficacy coefficients have the required effects directly on the decision-making related to the selection of handover.

As it was suggested, in the assessment of the functional pattern of the proposed scheme, using the concepts of Big Data is proposed as our leading approach toward achieving optimization in the network. The effect of some of the intended parameters is emerged as the determination of a threshold limit in the network. Also the effect of some of these parameters can be analyzed in a Fuzzy way as Fig. 2.

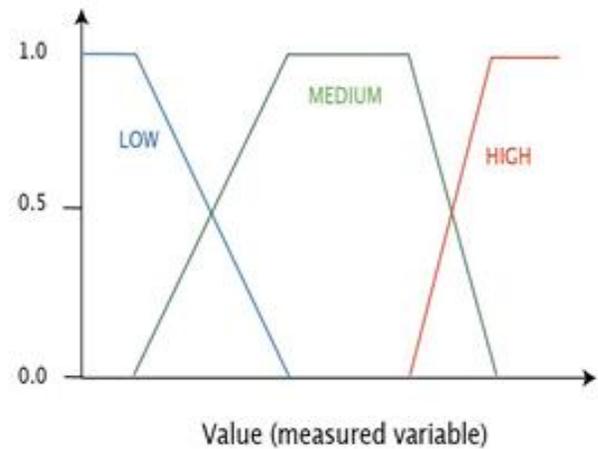

Fig. 2. A typical fuzzy values.

This decision-making pattern (NFDE) for the selection of an optimal network is a multi-dimensional, approximate and indefinite logic problem which is an appropriate candidate to be solved based on Fuzzy Nervous method. NFDE is expanded exactly based on Adaptive Fuzzy Logic System (AFLS). AFLS is a type of Fuzzy Logic System (FLS) that has fuzzifying and defuzzifying units. Its structure is similar to the traditional FLS but its rules are derived and extracted from a known educative data. In other words, its parameters can be educated just like in a nervous network method but along with its structure in a Fuzzy Logic System Structure.

Central defuzzifier is a very famous method for defuzzification but can't be used in NFDE by its calculated expense and prevents from using back-propagation education algorithm. The proposed AFLs include an alternative defuzzification method based on the FALCON model shown by Altug. NFDE feed-forward structure based on FALCON. NFDE is created when 1) the mobile host identifies and then specifies a new radio link, 2) the priority of the user changes, 3) a request for a new service is made and 4) there is a declination in the new signal or a complete signal loss from the current radio link.

Definition of membership functions: in choosing the best network in a change of vertical signal transfer, applying an appropriate parameter from different layers both on the user-side and on the system-side is required. The used features for NFDE includes financial costs (C), network bandwidth (B), RSS (R), user priority (U) and network delay (D). As a result, the revelation of a better control is possible by increasing the number if Fuzzy sets but this issue will reinforce the complexity of NFDE. Membership functions must be adaptable toward environment change in order to maintain its usefulness. The system behavior is relied on Fuzzy rules and membership functions significantly so as to describe decision-making rules. NFDE features are defined as the following.

## V. EVALUATION RESULTS

We have had an experimental investigation about the differences WLAN and UMTS based on their functionality per predefined constant values for some effective parameters in which, Cost, Bandwidth, Transmission Rate and Delay have an impressive role in this scenario.

As it can be seen in the result in the Table 3, the differences will be deeply dependent to the coefficient of these parameters.

TABLE III. THE IMPACT OF INDICATORS IN PHAZIZATION

| | C(Cent/Kb) | B (Mbit/s) | R (dBm) | D (ms) |
|---|---|---|---|---|
| **WLAN** | 0.001 | 11 | +38 | **1.25** |
| **UMTS** | 0.220 | 0.5 | -100 | **18.54** |

By doing repetitive simulation, below vector has been achieved as coefficient which can be used in next simulations.

*W= [0.0625, 0,0791, 0.0211, 0.0981, 0.4991]*

By considering the achieved coefficient for various parameters which are used in simulation, we arranged a wide evaluation for these two networks in which the combination of these two networks considered as a heterogeneous network. As Table 4 has shown, the functionality of the provided network has been related to the kind of service and the application type.

TABLE IV. THE SIMULATION RESULTS OF THE PROPOSED MODEL

| Network | Application Type | Service 1 | Service 1 |
|---|---|---|---|
| **WLAN** | Min | 0.775 | **0.011** |
| | Max | 3.675 | **0.055** |
| | Ave | 1.470 | **0.029** |
| **UMTS** | Min | 6.944 | **0.900** |
| | Max | 68.690 | **1.699** |
| | Ave | 19.721 | **1.191** |

In order to assess the results of suggested plan, we will try to analyse the output of performing suggested model in a cluster with five base station and 15 cells in KPI mode. Due to a lot of population concentration in two half of the year, this cluster can, as a suitable field for surveying the output of suggested model, be considered. By surveying KPI in primary half that the concentration of population is high, there has been observed this network from the attitude rate of sources and efficiency is in a suitable condition. Now, in the second half that the rate of population concentration is decreasing, if the sources of network are used like past, the rate of the efficiency of the network will be criticized. In these conditions, the presented optimized model, in order to devoting signalling sources to the traffic sources and increasing rate of transferring data, will have an approach based on decreasing the rate of signalling load. Due to this reason, by decreasing handover attempt rate, practically the rate of frequency interferences have been decreased and the quality of channels will increase, and anymore there is no need to create more signalling for creating new neighboring relations. For this reason, by hardening the threshold of handover, we possibly decrease the rate of signalling relations related to transferring service from one site to another site.

Mentioning this point is necessary that the number of users to be high, there are need for so many handover in the network; because the traffic load of network must be arranged based on the rate of present sources in every point. But in a condition that the number of users to be down, every cell can, in most cases, prepare its needed sources under its zone, except fluid users that in most cases of servicing to them, is along with handovers.

As it is observed in Fig. 3, with increasing the threshold of decision making for handover, the rate of handover attempt will be decreased noticeably. In continuance of effects, this change will be surveyed in the levels of other parameters KPI.

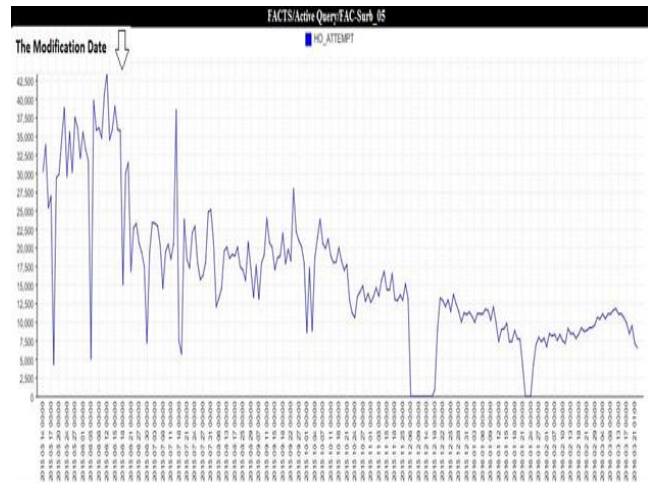

Fig. 3. Decreasing the rate of handover attempt in period of assessing cluster, before and after doing changes.

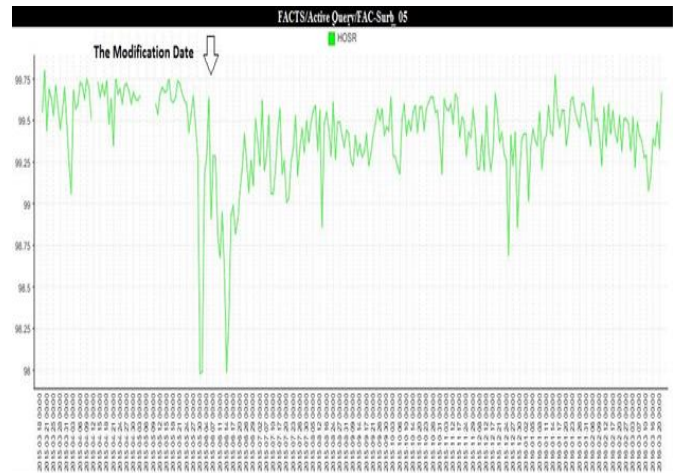

Fig. 4. The handover success rate before and after doing modifications.

With regard to this point that the number of demands have been decreased noticeably for handover, but, successful rate of handover (HOSR) only has been decreased a few percent. As it is observed in Fig. 4, this rate of reduction, will not have negative effect on the efficiency of the network. And but, main effectiveness of hardening conditions of handover is in reducing the load of signalling in the network. It is clear in Fig. 5 that after doing this changing in network, the load of signalling has had reduction % 30 that this case, for getting to an optimized network can be very significant. With regard to these conditions, we can devote unusable sources of network signalling to traffic channels for transferring data.

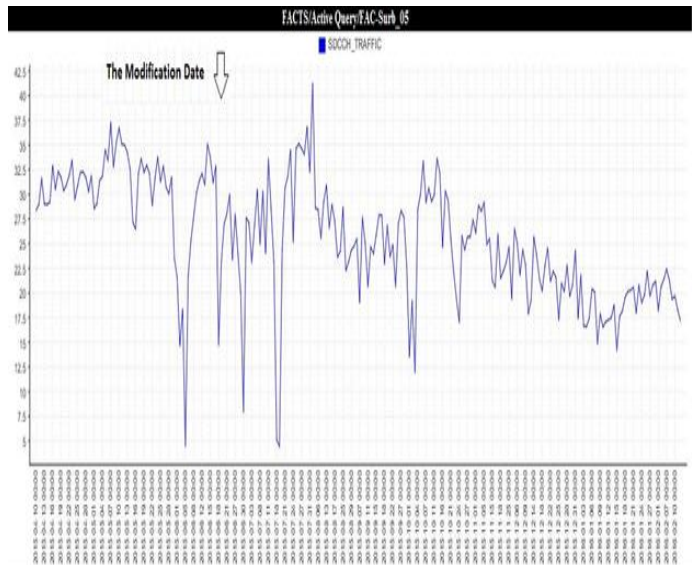

Fig. 5. 30 percent reduction of average rate of traffic load in signalling channels for every site of cluster.

## VI. CONCLUSION AND FUTURE WORK

In this paper, a new method of Joint Radio Resource Management was proposed which aside from providing a desirable service quality, decreases the vertical handover rate and service cost. In the considered scenario, both the immediate and non-immediate services are considered. In this paper, we tried to deal with optimization of traffic distribution in heterogeneous networks on two levels. On the first level, this issue will be done by selecting an appropriate RAT and on the second level, by establishing the user connection to the specified Femtocell's cells during the process of optimization. Using Big Data technique will have tremendous effects on prediction feasibility and better future decision-making about network load distribution which this decision-making will be done during user-level and network-level processing.

Future work could include extending the proposed method for other application such VANET. Another area of research related to this research is the adaptation of proposed method for distributed and cloud environments.

REFERENCES

[1] L. Giupponi, R. Agusti, J. Perez-Romero and O. Sallent. "Joint Radio Resource Management Algorithm for Multi-RAT Networks," IEEE Global Telecommunications Conference, GLOBECOM '05, pp. 3850-3855, Dec. 2005.

[2] 3GPP, "Improvement of RRM across RNS and RNS/BSS (post rel-5) (release 6)," TR 25.891 v0.3.0, 3rd Generation Partnership Project (3GPP), 2003.

[3] D. Niyato and E. Hossain. "Wlc04-5: Bandwidth allocation in 4g heterogeneous wireless access networks: A noncooperative game theoretical approach," IEEE Global Telecommunications Conference, GLOBECOM '06, pp. 1–5, Nov. 2006.

[4] A. Tolli, P. Hakalin, "Adaptive load balancing between multiple cell layers," VTC 2002-Fall. 2002 IEEE 56th Vol. 3, pp. 1691–1695, Sept. 2002.

[5] K. Murray, R. Mathur, D. Pesch, "Network access and handover control in heterogeneous wireless networks for smart space environments, " in: First International Workshop on Management of Ubiquitous Communications and Services, MUCS, Waterford, Ireland, Dec. 2003.

[6] W. Zhang, "Performance of real-time and data traffic in heterogeneous overlay wireless networks, " in: Proceedings of the 19th International Teletraffic Congress (ITC 19), Beijing, 2005.

[7] E. Vanem, S. Svaet and F. Paint. "Effects of multiple access alternatives in heterogeneous wireless networks," in Proc. IEEE WCNC, pp. 1696–1700, 2003.

[8] J. P. Romero, O. Sallent, R. Agusti, and M. A. Diaz-Guerra, Radio Resource Management Strategies in UMTS: John Wiley & Sons, 2005.

[9] Xu Yang, J. Bigham, and L. Cuthbert. "Resource management for service providers in heterogeneous wireless networks," Wireless Communications and Networking Conference, IEEE, Vol. 3, pp. 1305–1310, Mar. 2005.

[10] F. IdrisKhan and E. Huh. "An adaptive resource management for mobile terminals on vertical handoff," Annals of Telecommunications, pp. 435-447, May. 2008.

[11] K. Piamrat, C.Viho, A. Ksentini and J.M. Bonnin. "Resource Management in Mobile Heterogeneous Networks: State of the Art and Challenges," inria-00258507, v 4-3, Mar. 2008.

[12] G. Koundourakis, D.I. Axiotis, and M. Theologou. "Network-based access selection in composite radio environments," Wireless Communications and Networking Conference, 2007.WCNC 2007. IEEE, pp 3877–3883, Mar. 2007.

[13] O. Ormond, P. Perry, and J. Murphy. "Network selection decision in wireless heterogeneous networks," Personal, Indoor and Mobile Radio Commu- nications, 2005. PIMRC 2005. IEEE 16th International Symposium on, Vol. 4, pp. 2680–2684, Sept. 2005.

[14] Xiaoshan Liu, Victor O. K. Li, and Ping Zhang. "NXG04-4: Joint radio resource management through vertical handoffs in 4g networks," IEEE Global Telecommunications Conference, 2006. GLOBECOM '06, pp. 1–5, Nov. 2006.

[15] Kalaiselvi, D., and R. Radhakrishnan. "Multiconstrained QoS routing using a differentially guided krill herd algorithm in mobile ad hoc networks."Mathematical Problems in Engineering 2015 (2015).

[16] Xi, Yufang, and Edmund M. Yeh. "Distributed algorithms for minimum cost multicast with network coding." IEEE/ACM Transactions on Networking 18, no. 2 (2010): 379-392.

[17] Alam, Muhammad, Du Yang, Kazi Huq, Firooz Saghezchi, Shahid Mumtaz, and Jonathan Rodriguez. "Towards 5G: Context Aware Resource Allocation for Energy Saving." Journal of Signal Processing Systems 83, no. 2 (2016): 279-291.

[18] Foster, Gerry, Seiamak Vahid, and Rahim Tafazolli. "SON Evolution for 5G Mobile Networks." Fundamentals of 5G Mobile Networks (2015): 221-240.

[19] Imran, Ali, Ahmed Zoha, and Adnan Abu-Dayya. "Challenges in 5G: how to empower SON with big data for enabling 5G." IEEE Network 28, no. 6 (2014): 27-33.

[20] Murugeswari, R., S. Radhakrishnan, and D. Devaraj. "A multi-objective evolutionary algorithm based QoS routing in wireless mesh networks."Applied Soft Computing 40 (2016): 517-525.

[21] Fadlullah, Zubair Md, Duong Minh Quan, Nei Kato, and Ivan Stojmenovic. "GTES: An optimized game-theoretic demand-side management scheme for smart grid." IEEE Systems journal 8, no. 2 (2014): 588-597.

[22] Liu, Yi, Chau Yuen, Shisheng Huang, Naveed Ul Hassan, Xiumin Wang, and Shengli Xie. "Peak-to-Average Ratio Constrained Demand-Side Management with Consumer's Preference in Residential Smart Grid." IEEE Journal of Selected Topics in Signal Processing 8, no. 6 (2014): 1084-1097.